\documentclass{elsart3p}

\usepackage{graphicx}
\usepackage{amssymb}

\begin{document}

\begin{frontmatter}

%titles, authors and addresses

%\title{Title\thanksref{label1}}

\title{Comparative investigation of the coupled-tetrahedra quantum
 spin systems Cu$_2$Te$_2$O$_5$X$_2$, X=Cl, Br ~ and  Cu$_{4}$Te$_{5}$O$_{12}$Cl$_{4}$}


\author[a]{R. Valent{{\'\i}}}
\author[b]{T. Saha-Dasgupta}
\author[a]{H. O. Jeschke}
\author[b]{B. Rahaman}
\author[c]{H. Rosner}
\author[d]{P. Lemmens}
\author[e]{R. Takagi}
\author[e]{M. Johnsson}

\address[a]{Institut f{{\"u}}r Theoretische Physik, Universit{{\"a}}t Frankfurt,
Max-von-Laue-Str. 1, 60438 Frankfurt, Germany}
\address[b]{S.N. Bose National Centre for Basic Sciences,
JD Block, Sector 3,
 Salt Lake City, Kolkata 700098, India.}
\address[c]{MPI-CPS, Noethnitzer Str. 40, 01187 Dresden, Germany.}
\address[d]{IPCM, TU Braunschweig, Mendelssohnstrasse, 3, 38106
Braunschweig, Germany.}
\address[e]{Department of Inorganic Chemistry, Stockholm University,
S-106 91 Stockholm, Sweden.}

\begin{abstract}

We present a comparative study of the coupled-tetrahedra quantum spin systems
Cu$_2$Te$_2$O$_5$X$_2$, X=Cl, Br (Cu-2252(X)) and the newly synthesized 
Cu$_{4}$Te$_{5}$O$_{12}$Cl$_{4}$ (Cu-45124(Cl)) based on {\it ab initio} Density
Functional Theory calculations. The magnetic behavior of Cu-45124(Cl) with a phase
transition to an ordered state at a lower critical temperature T$_c$=13.6K  than in
Cu-2252(Cl) (T$_c$=18K) can be well understood in terms of the modified interaction
paths. We identify the relevant structural changes between the two systems and discuss
the hypothetical behavior of the not yet synthesized Cu-45124(Br) with an {\it ab
initio} relaxed structure using Car-Parrinello Molecular Dynamics.
\end{abstract}

\end{frontmatter}

% main text
%\section{Introduction}
\label{Introduction} Frustrated quantum spin systems have received considerable
attention in recent years due to their unconventional behavior. One class of materials
belonging to this family are the  oxohalogenides Cu$_2$Te$_2$O$_5$X$_2$ (Cu-2252(X)),
X=Cl, Br \cite{Johnsson00} which contain weakly coupled Cu$_4$$^{2+}$ tetrahedra. These
systems show incommensurate long-range ordering  at temperatures $T_c$= 18.2~K
(Cu-2252(Cl)) and $T_c$=11.4~K (Cu-2252(Br)) \cite{Lemmens01,Zaharko04}  with strongly
reduced ordered moments for the latter system. Evidence for low energy longitudinal
magnon excitations have been reported for Cu-2252(Br) \cite{Gros03} and interpreted as a
manifestation of the proximity of this system to quantum criticality. An {\it
ab initio} study on the nature of the microscopic interactions
 performed by us \cite{Valenti03} revealed
that the knowledge of the subtle ratio between intra-tetrahedra and inter-tetrahedra
couplings is crucial for understanding the behavior of these materials. To shed
further light on the physics of this interesting class of materials we add the recently
discovered oxohalogenide Cu$_{4}$Te$_{5}$O$_{12}$Cl$_{4}$ (Cu-45124(Cl))
\cite{Takagi:06} to our considerations and compare it with the previously known Cu-2252.

%\section{Experimental}
\label{Experimental} Magnetic susceptibility experiments performed on the novel
Cu-45124(Cl) system show for temperatures below approximately 30~K a different evolution
of the antiferromagnetic correlations \cite{Takagi:06}, compared to the Cu-2252 systems.
 This can be described by a shift
of the maximum in the susceptibility from 19~K for Cu-45124(Cl) to 23~K for Cu-2252(Cl)
and even to 30~K for Cu-2252(Br). From this shift we derive a smaller averaged magnetic
energy scale of the novel system. A kink in the susceptibility marks long range ordering
in Cu-45124(Cl) at T$_{c}$=13.6~K. In the other two compounds the change of slope is
weaker and observed at 18.2 and 11.4~K, respectively. Raman scattering experiments show
a doublet of sharp excitations with a mean energy E$_{m}$=36.9~cm$^{-1}$. This
energy scale is also reduced compared to the reference systems with 
 E$_{m}$=47.5 and 60~cm$^{-1}$ for Cu-2252(X) X=Cl and  Br respectively.
 It is noteworthy that the
energy of this scattering scales well with the maximum temperature of the
susceptibility.

%\appendix

\begin{figure}[h]
\includegraphics[height=0.4\textwidth,angle=-90,keepaspectratio]{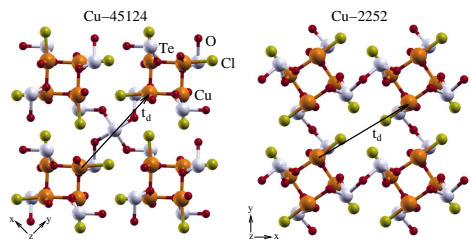}
\caption{ Crystal structure of  Cu-45124(Cl) (left panel) and
Cu-2252(Cl) (rigth panel)  projected on the $ab$ plane.}

\label{struc1}
\end{figure}

%\section{Microscopic model}
\label{Microscopic model} The structural difference between Cu-45124(Cl)
\cite{Takagi:06} and Cu-2252(Cl) \cite{Johnsson00} is given by an additional TeO$_4$
complex in the middle of the Cu-tetrahedral network in the $ab$ plane, increasing the
distance between the Cu$_4$ clusters with respect to Cu-2252(Cl) (see Fig. \ref{struc1})
and the stacking of the Cu$_4$ tetrahedra along the c-axis alternates between successive
rows \cite{Takagi:06}.
This results in a centrosymmetric 
$P4/n$ structure for Cu-45124(Cl) in comparison to the noncentrosymmetric  $P\bar{4}$ for
 Cu-2252(Cl).

We performed {\it ab initio} DFT calculations and analyzed the resulting
electronic structure in terms of the  NMTO-downfolding
technique \cite{nmto} in order to investigate the various interaction paths in this
system and the relative importance of the intra- to inter-tetrahedral couplings. In Ref. \cite{Rahaman:06} we give a detailed account of the
various hopping terms. The most remarkable change observed in this new material is the
drastic reduction of the in-plane inter-tetrahedral diagonal interaction $t_d$ (see Fig.
\ref{struc1}) which was shown  \cite{Valenti03}  to be mediated by the Cl $p$ orbitals
and has been pointed out to play an important role in explaining the magnetic properties
\cite{Valenti03,Zaharko06} of the Cu-2252 compounds. This reduction is based on the different
alignment of the Cu-Cl bonds belonging to neighboring Cu$_4$ tetrahedra which are
parallel to each other in Cu-45124(Cl) instead of pointing to each other as it is the
case in Cu-2252(Cl). The resulting Cl$p$-Cl$p$ $\pi$ type bonding in Cu-45124(Cl) is
much weaker than the Cl$p$-Cl$p$ $\sigma$ bonding in Cu-2252(Cl).
% In Fig. \ref{Wannier}
%we show the Cu-d$_{x^2-y^2}$ downfolded NMTO Wannier orbitals placed at two Cu sites
%situated at the in-plane diagonal positions. 
In this sense, this new compound belongs to
the limit of very weakly coupled Cu$_4$-tetrahedron systems. This feature may
explain the magnetic ordering at a lower temperature than in the Cu-2252(Cl) system.

%\begin{figure}[h]
%\includegraphics[width=6cm,keepaspectratio]{diagonal.eps}
%\caption{Cu-d$_{x^2-y^2}$ downfolded NMTO Wannier orbitals }

%\label{Wannier}
%\end{figure}

In an attempt to predict the behavior of the not yet synthesized Cu-45124(Br) and
motivated by the more anomalous properties of Cu-2252(Br) compared to Cu-2252(Cl), we have
performed a geometry relaxation for Cu-45124(Br) in the framework of {\it ab initio}
molecular dynamics \cite{AIMD,PAW} and analyzed the electronic properties with the
NMTO-downfolding technique. While the details of these calculations are presented
elsewhere \cite{Rahaman:06}, we observe that the intra-tetrahedron frustration changes
 in this system with respect to Cu-45124(Cl) and Cu-2252(Br)
 while the inter-tetrahedra coupling
remains weak. An experimental investigation of this not yet synthesized compound would
therefore be highly desirable in the context of frustrated cluster magnetism.

%\section{Discussion and conclusions}
\label{Discussion and conclusions}
% Discussion on the previous points

{\bf  Acknowledgements:} We thank W. Brenig, V. Gnezdilov, and R.K. Kremer for important
discussions. This work has been supported by the DFG, the Emmy-Noether Program,
the ESF-HFM, the MPG-Indian
partnergroup and the Swarnajaynati Fellowship.

%\acknowledgements

\end{document}